\documentclass[]{svjour2}

\smartqed

\usepackage{amssymb}
\usepackage{graphicx}

\journalname{J Stat Phys}

\begin{document}

\title{Counter-ions between or at asymmetrically charged walls: \\
2D free-fermion point}

\titlerunning{Counter-ions between or at asymmetrically charged walls}

\author{Ladislav \v{S}amaj \and Emmanuel Trizac} 

\institute{Ladislav \v{S}amaj \and Emmanuel Trizac \at
Universit\'e Paris-Sud, Laboratoire de Physique Th\'eorique et 
Mod\`eles Statistiques, UMR CNRS 8626, 91405 Orsay, France \and
Ladislav \v{S}amaj \at Institute of Physics, Slovak Academy of Sciences,
D\'ubravsk\'a cesta 9, 845 11 Bratislava, Slovakia}

\date{Received: / Accepted: }

\maketitle

\begin{abstract}
This work contributes to the problem of determining effective 
interaction between {\em asymmetrically} (likely or oppositely) charged 
objects whose total charge is neutralized by mobile pointlike counter-ions 
of the same charge, the whole system being in thermal equilibrium.
The problem is formulated in two spatial dimensions with logarithmic Coulomb
interactions. 
The charged objects correspond to two parallel lines at distance $d$, with 
fixed line charge densities.
Two versions of the model are considered: the standard ``unconstrained''
one with particles moving freely between the lines and the ``constrained''
one with particles confined to the lines.
We solve exactly both systems at the free-fermion coupling and compare 
the results for the pressure (i.e. the force between the lines per unit 
length of one of the lines) with the mean-field Poisson-Boltzmann solution.
For the unconstrained model, the large-$d$ asymptotic behaviour of
the free-fermion pressure differs from that predicted by the mean-field theory.
For the constrained model, the asymptotic pressure coincides with 
the attractive van der Waals-Casimir fluctuational force.
For both models, there are fundamental differences between the cases of
likely-charged and oppositely-charged lines, the latter case corresponding 
at large distances $d$ to a capacitor.


\end{abstract} 

\section{Introduction} \label{sec:1}
We study classical (i.e. non-quantum) systems of particles interacting 
pairwisely by the Coulomb potential, which are in thermal equilibrium.
Such systems are of practical interest in the ``real'' three-dimensional (3D)
world where the Coulomb potential has the standard $1/r$ form.
One can extend its definition to any dimension $\nu=1,2,\ldots$ in the
following way: the Coulomb potential $v$ at a spatial position 
${\bf r}\in \mathbb{R}^{\nu}$, induced by a unit charge at the origin 
${\bf 0}$, is the solution of the Poisson equation
\begin{equation}
\Delta v({\bf r}) = - s_{\nu} \delta({\bf r}) ,
\end{equation}
where $s_{\nu}=2\pi^{\nu/2}/\Gamma(\nu/2)$ [$\Gamma(x)$ denotes the Gamma 
function] is the surface area of the $\nu$-dimensional unit sphere.
In particular,
\begin{equation}
v({\bf r}) = \left\{
\begin{array}{ll}
-\ln(r/r_0) \phantom{aaa} & \mbox{if $\nu=2$,} \cr & \cr
\displaystyle{\frac{r^{2-\nu}}{\nu-2}} & \mbox{otherwise,} 
\end{array} \right.
\end{equation}
where $r\equiv \vert {\bf r}\vert$ and an arbitrary length scale $r_0$ 
fixes the zero of the two-dimensional (2D) Coulomb potential.
Such definition implies in the Fourier space the characteristic $1/k^2$
behaviour which maintains many generic properties of 3D Coulomb systems
like screening \cite{Martin88}. 
This is why many important phenomena related to the Coulomb law
manifest themselves in a similar way in various spatial dimensions.

2D one-component (jellium) and symmetric two-component (Coulomb gas) systems 
with logarithmic charge interactions are of special importance because they 
are exactly solvable, besides the mean-field weak-coupling limit, 
also at a specific finite temperature. 
The solvable cases involve the bulk regime \cite{Jancovici81,Cornu87} as well 
as inhomogeneous, semi-infinite or fully finite geometries \cite{Cornu89}, 
see reviews \cite{Jancovici92,Forrester98}.
This permits us to check basic concepts or ideas on 2D exact solutions.

One of extensively studied problems in our days is the electromagnetic 
Casimir effect, see e.g. \cite{Bordag01} and \cite{Milton01}.
In its formulation within classical statistical mechanics, two conducting 
neutral slabs in thermal equilibrium [at the inverse temperature 
$\beta=1/(k_{\rm B}T)$], being from one another at distance $d$ much larger 
than any microscopic (e.g. Debye-H\"uckel) length scale, are attracted by 
the long-range van der Waals-Casimir force.
This force is due to thermal fluctuations of local charge densities
inside globally neutral conductors.
In $\nu$ spatial dimensions, the force per unit area of one of the slabs,
or equivalently the pressure $P$, was obtained in the form \cite{Jancovici04}
\begin{equation} \label{Casimir}
\beta P = - \frac{(\nu-1)\zeta(\nu)\Gamma(\nu/2)}{2^{\nu}\pi^{\nu/2}} 
\frac{1}{d^{\nu}} \qquad \mbox{(dimension $\nu$),}
\end{equation}
where $\zeta(\nu) = \sum_{n=1}^{\infty} n^{-\nu}$ is the Riemann zeta function
and the minus/plus sign means attraction/repulsion of the slabs.
Note that in one dimension, one gets a finite result owing to
the limiting behaviour $\lim_{\nu\to 1} (\nu-1) \zeta(\nu) = 1$.
In 2D, we have explicitly
\begin{equation} \label{Casimir2D}
\beta P = - \frac{\pi}{24} \frac{1}{d^2} \qquad \mbox{(2D)}, 
\end{equation}
and in 3D
\begin{equation} \label{Casimir3D}
\beta P = - \frac{\zeta(3)}{8\pi} \frac{1}{d^3} \qquad \mbox{(3D)} .
\end{equation}
These large-distance behaviours do not depend on the composition of
conducting slabs, nor on the temperature: they are universal.
It was shown \cite{Buenzli05} that the microscopic origin of this universality 
lies in an electroneutrality sum rule for Coulomb fluids. 

Another basic problem in soft matter physics is the determination of the 
temperature-dependent effective interaction between charged mesoscopic objects, 
macro-ions (colloids, polyelectrolytes) or charged membranes of several 
thousands elementary charges $e$, immersed in a polar solvent, e.g. water, 
containing mobile micro-ions of low valence.
Let the large macro-ions be approximated by rectilinear hard walls, filled
by a material of the same dielectric constant as the solvent (i.e. there are
no electrostatic image charges). 
In a simplified model coined as ``counter-ions only'', or ``salt-free'', 
the macro-ions acquire 
a fixed surface charge density $\sigma e$ ($\sigma>0$) by releasing equally 
charged $-e$ (valence $=1$, for simplicity) counter-ions into the solvent, 
the system as a whole being electroneutral.
This model was studied in the semi-infinite geometry of one charged wall and 
for counter-ions being in between two {\em symmetrically} charged parallel
plates at distance $d$.
It became the cornerstone for studying systematically the weak-coupling 
(high-temperature) \cite{Netz00} and strong-coupling
(low-tem\-pera\-ture) \cite{Netz01,Samaj11a} limit of Coulomb fluids.
While two symmetrically charged plates always repel one another in the
weak-coupling limit, the situation changes for sufficiently large couplings: 
an attraction may occur in a certain interval of distances $d$. 
An important information in the two-plate geometry is the
asymptotic $d\to\infty$ behaviour of the pressure, namely its sign
(attraction or repulsion) and the explicit inverse-power-law dependence on $d$.
Naively, one may expect that for large $d$, the counter-ions will separate
onto two sets, one for each plate, to neutralize these plates.
Then each plate with its counter-ions might be viewed as a ``neutral 
slab'' and the two slabs interact consequently by the attractive van der 
Waals-Casimir force of type (\ref{Casimir}).  
But the Poisson-Boltzmann solution \cite{Andelman06}, generalized 
straightforwardly from 3D to any dimension $\nu$,
\begin{equation} \label{pressureas}
\beta P \mathop{\sim}_{\sigma d\to\infty} \frac{2\pi^2}{\beta e^2 s_{\nu}} 
\frac{1}{d^2} \qquad \mbox{(dimension $\nu$)}
\end{equation}
shows that the possibility of counter-ions to migrate in the space between 
the plates is crucial.
This force is repulsive and differs fundamentally from the attractive 
fluctuational one (\ref{Casimir}): although the prefactor to $1/d^2$ 
is independent of the fixed surface charge density $\sigma e$, 
it does depend on the temperature.
An important question is whether the mean-field asymptotic behaviour 
(\ref{pressureas}) remains valid also for finite temperatures.
Such supposition is often taken as natural, without any rigorous proof.

For the related geometry of one wall, the density profiles of counter-ions
were investigated within a test-charge theory for the electric double layer
in ref. \cite{Burak04}.
This theory provides results which interpolate between the correct 
strong-coupling (exponential decay) and the weak-coupling (algebraic decay) 
limits.
Its application to intermediate values of the coupling constant indicates 
that there is a crossover from exponential to algebraic decay at large 
distance from the charged wall.
It is argued that in the algebraic regime, the density profile is described
by a modified mean-field equation.
This unexpected conclusion might be put into doubts by the fact that the
exact contact theorem for the particle density at the wall is not fulfilled 
by the test-charge theory in the region of intermediate values of 
the coupling constant.

For symmetrically charged lines in 2D, the mean-field formula for 
the pressure (\ref{pressureas}) takes the form
\begin{equation} \label{meanfield}
\beta P \mathop{\sim}_{\sigma d\to\infty} \frac{\pi}{\Gamma} \frac{1}{d^2} 
\qquad (\Gamma\to 0),
\end{equation}
where $\Gamma\equiv \beta e^2$ is the 2D coupling constant.
In a recent work \cite{Samaj11b}, we solved exactly the free-fermion point
$\Gamma=2$ by using a technique of anticommuting variables, with the result
\begin{equation} \label{Gamma2}
\beta P = \frac{1}{\pi d^2} \int_0^{2\pi\sigma d} {\rm d}s\, \frac{s}{\sinh s}
{\rm e}^{-s} \mathop{\sim}_{\sigma d\to\infty} \frac{\pi}{12} \frac{1}{d^2}
\qquad (\Gamma=2).
\end{equation} 
This asymptotic behaviour is not compatible with the mean-field one 
(\ref{meanfield}), i.e. the mean-field prefactor must be renormalized 
by a $\Gamma$-dependent function which goes to unity as $\Gamma\to 0$.

\begin{figure}[htb]
\begin{center}
\includegraphics[width=0.6\textwidth,clip]{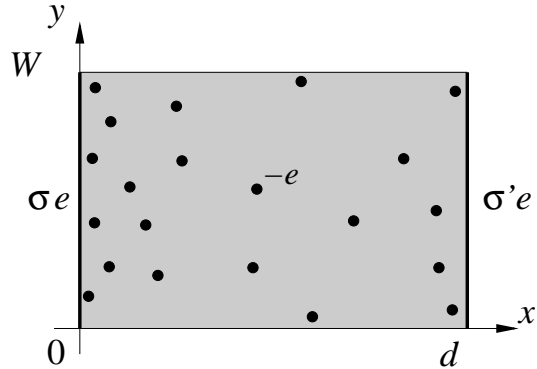}
\caption{Cylinder geometry with periodical boundary conditions (period $W$)
along the $y$-axis. 
Two parallel lines with  fixed charge densities $\sigma e$ and $\sigma' e$ 
are localized at the end points $x=0$ and $x=d$, respectively. 
Pointlike counter-ions of charge $-e$ are allowed to move freely between 
the lines in the ``unconstrained'' version of the model.}
\label{fig:1}
\end{center}
\end{figure}

\begin{figure}[htb]
\begin{center}
\includegraphics[width=0.6\textwidth,clip]{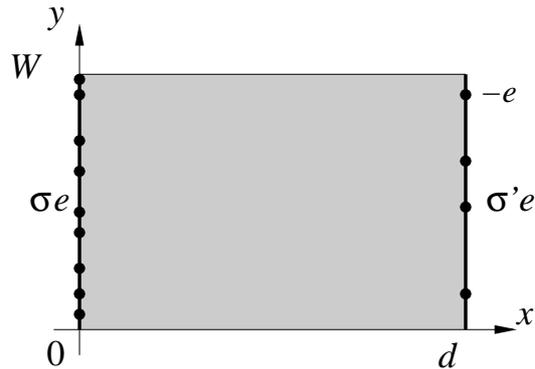}
\caption{``Constrained'' version of the model: Cylinder geometry with 
asymmetrically charged lines, where counter-ions' displacements are limited to 
the lines.}
\label{fig:2}
\end{center}
\end{figure}

The aim of this paper is twofold.
\begin{itemize}
\item
First, for the ``unconstrained'' version of the model with counter-ions 
moving freely in space between the lines, we generalize the exact result at 
$\Gamma=2$ for symmetric lines \cite{Samaj11b} to asymmetrically charged lines, 
see Fig. \ref{fig:1}.
The weak-coupling and strong-coupling limits for 3D asymmetrically
charged plates were investigated in ref. \cite{Kanduc08}.
The 2D exact result at the finite free-fermion coupling $\Gamma=2$,
presented in this work, leads to a pressure which consists of two decoupled
contributions from each of the lines. 
\item
Our second aim is to solve a ``constrained'' version of the model in which 
counter-ions do not move freely between the asymmetrically charged lines, 
but are constrained to them (Fig. \ref{fig:2}). 
The particle occupation of the lines is not fixed, i.e. the particles
can lie either on $x=0$ line or $x=d$ line. 
A similar 3D model, in which charged particles are constrained to and 
simultaneously neutralize each of the symmetrically-charged plates, 
was introduced and studied at low temperatures in refs. \cite{Lau00,Lau01}.  
We show that our constrained model yields a phenomenology that is close to 
the Casimir electromagnetic effect, although, at each finite distance, 
the net charge of every line is non-zero and thus the charged lines together 
with the attached particles do not describe neutral conductors as required 
by the Casimir effect. 
\end{itemize}
For both models, there are important differences between the cases of
likely-charged and oppositely-charged lines, the latter case corresponds 
at large distances to a capacitor.

The paper is organized as follows.
In Sect. 2, we recapitulate briefly the formalism of anticommuting variables 
for a cylinder geometry of the confining domain, with two asymmetrically 
charged circles at distance $d$.
The thermodynamic limit is reached by increasing the cylinder circumference
$W$ to infinity, keeping the overall electroneutrality, which effectively 
transforms discrete sums to continuous integrals.
Sect. \ref{sec:3} is devoted to the solution of the unconstrained model.
In Sect. \ref{sec:4}, we deal with the model of counter-ions constrained to 
the charged lines for which a comparison is made for large distances $d$ 
with the van der Waals-Casimir asymptotic formula (\ref{Casimir2D}).
The qualitatively distinct regimes with likely-charged and oppositely-charged
lines are analyzed for both models.
A short summary and conclusions are drawn in Sect. \ref{sec:concl}.

\section{General formalism for cylinder geometry} 
\label{sec:2}
We consider the system of $N$ mobile pointlike particles with the (minus) 
elementary charge $-e$, confined to the surface of a cylinder of 
circumference $W$ and finite length $d$.
The surface of the cylinder can be represented as a 2D semiperiodic rectangle
domain $\Lambda$ of points ${\bf r}=(x,y)$ with coordinates 
$x\in [0,d]$ (free boundary conditions at $x=0,d$) and
$y\in [0,W]$ (periodic boundary conditions at $y=0,W$).
It is useful to introduce the complex coordinates $z=x+{\rm i}y$ and
$\bar{z}=x-{\rm i}y$.
There are fixed line charge densities $\sigma e$ and $\sigma' e$
of dimension [length]$^{-1}$ along the $y$-axis at the end-points 
$x=0$ and $x=d$, respectively.
Without any loss of generality, we restrict ourselves to
\begin{equation} \label{restriction}
\sigma>0 , \qquad -\sigma < \sigma' \le \sigma ,
\end{equation}
which therefore includes cases where $\sigma'<0$.
We introduce the asymmetry parameter
\begin{equation}
\eta \equiv \frac{\sigma'}{\sigma} , \qquad \eta\in (-1,1] .
\end{equation}
The symmetric case corresponds to $\eta=1$.
The overall electroneutrality condition is expressed as
\begin{equation}
N = (\sigma+\sigma') W .
\end{equation} 
Note that the requirement of positivity of $N$ corresponds to the
restriction $-\sigma<\sigma'$ in (\ref{restriction}). 
We are interested in the thermodynamic limit $N,W\to\infty$, 
keeping the ratio $N/W=\sigma+\sigma'$ fixed.

The Coulomb potential $v$ at a spatial position ${\bf r}\in\Lambda$, induced
by a unit charge at the origin ${\bf 0}$, is defined as the solution of 
the 2D Poisson equation $\Delta v({\bf r}) = -2\pi\delta({\bf r})$, under
the periodicity requirement along the $y$-axis with period $W$.
Considering the potential as a Fourier series in $y$, one obtains
\cite{Choquard81}  
\begin{eqnarray}
v({\bf r}) & = & - \ln \left\vert 2 \sinh\left( \frac{\pi z}{W} \right)
\right\vert \nonumber \\ & = & - \frac{1}{2}
\ln \left[ 2 \cosh\left( \frac{2\pi x}{W} \right) - 
2 \cos\left( \frac{2\pi y}{W} \right) \right] .
\end{eqnarray} 
For small distances ${\bf r}\ll W$, this potential reduces to the 2D
Coulomb potential $-\ln(2\pi r/W)$.
At large distances along the cylinder $x\gg W$, this potential behaves like
the 1D Coulomb potential $-\pi \vert x\vert/W$.
The interaction potential of two unit charges at points ${\bf r}$ and
${\bf r}'$ is given by $v({\bf r},{\bf r}') = v(\vert {\bf r}-{\bf r}'\vert)$.
In what follows, we shall use two formulas:
\begin{equation} \label{formula1}
\int_0^W {\rm d} y\, v({\bf r}) = - \pi x 
\end{equation}
and
\begin{equation} \label{formula2}
\left\vert 2 \sinh \frac{\pi (z-z')}{W} \right\vert
= {\rm e}^{\frac{\pi}{W}(x+x')} \left\vert
{\rm e}^{-\frac{2\pi}{W}z} - {\rm e}^{-\frac{2\pi}{W}z'} \right\vert .
\end{equation}

For a given spatial configuration $\{ {\bf r}_1, \cdots, {\bf r}_N \}$
of $N$ charges, the total Coulomb energy of the system consists of the self
and mutual interactions of the fixed surface charge densities $\sigma e$ 
and $\sigma' e$, $E_{ss}$, particles and surface charge densities, 
$E_{ps}$, and particles themselves, $E_{pp}$.
Using the relation (\ref{formula1}) and the notation $v({\bf r})\equiv v(x,y)$,
the self-interaction of the surface charge densities $\sigma e$ and 
$\sigma' e$ vanishes,
\begin{equation}
\frac{1}{2} \left[ (\sigma e)^2 + (\sigma' e)^2 \right]
\int_0^W {\rm d}y \int_0^W {\rm d}y'\, v(0,y-y') = 0 ,
\end{equation}
while their mutual interaction energy is given by
\begin{equation}
(\sigma e) (\sigma' e) \int_0^W {\rm d}y \int_0^W {\rm d}y'\, v(d,y-y') 
= - \pi (\sigma e) (\sigma' e) W d .
\end{equation}
Since $(-e)(\sigma e)\int_0^W {\rm d}y\, v(x,y)=\pi\sigma e^2 x$, 
the interaction energy of particles with the surface charge densities 
takes the form
\begin{eqnarray}
E_{ps} & = & \sum_{j=1}^N \left[ \pi\sigma e^2 x_j + \pi\sigma' e^2 (d-x_j) 
\right] \nonumber \\ & = &
\sum_{j=1}^N \pi(\sigma-\sigma') e^2 x_j + N\pi\sigma' e^2 d .
\end{eqnarray}
The particle-particle interaction energy is simply given by
\begin{equation}
E_{pp} = \sum_{(j<k)=1}^N e^2 v(\vert {\bf r}_j-{\bf r}_k\vert) . 
\end{equation}
At inverse temperature $\beta=1/(k_{\rm B}T)$, the Boltzmann factor of
the total energy $E_N(\{{\bf r}\})=E_{ss}+E_{ps}+E_{pp}$ reads 
\begin{eqnarray}
{\rm e}^{-\beta E_N(\{{\bf r}\})} & = & {\rm e}^{-\pi\Gamma(\sigma')^2 W d}
\prod_{j=1}^N {\rm e}^{\pi\Gamma(\sigma'-\sigma)x_j} \nonumber \\ & & \times
\prod_{(j<k)=1}^N \left\vert 2 \sinh\frac{\pi(z_j-z_k)}{W} \right\vert^{\Gamma} ,
\end{eqnarray}
where $\Gamma=\beta e^2$ is the coupling constant.

Within the canonical ensemble, the partition function is defined as
\begin{equation} \label{partition}
Z_N = \frac{1}{N!} \int_{\Lambda} \frac{{\rm d}{\bf r}_1}{\lambda^2} 
\epsilon({\bf r}_1) \cdots \int_{\Lambda} \frac{{\rm d}{\bf r}_N}{\lambda^2} 
\epsilon({\bf r}_N) {\rm e}^{-\beta E_N(\{{\bf r}\})} ,
\end{equation}
where $\lambda$ is the thermal de Broglie wavelength and the function 
$\epsilon({\bf r})$ reflects a possible constraint to the location of particles.
In particular,
\begin{equation} \label{epsunconstrained}
\epsilon({\bf r}) = 1 
\end{equation}
for the unconstrained model pictured in Fig. \ref{fig:1} and
\begin{equation} \label{epsconstrained}
\epsilon({\bf r}) \equiv \epsilon(x) = 
\lambda \left[ \delta(x) + \delta(x-d) \right] 
\end{equation}
for the constrained model in Fig. \ref{fig:2}. 
Here, $\lambda$ is included for dimensional reason, and turns immaterial 
in what follows.
The introduction of the $\epsilon$-function allows us to treat the two models
using the same technique.
With the aid of formula (\ref{formula2}), the partition function 
(\ref{partition}) can be expressed as
\begin{equation} \label{partf}
Z_N = \left( \frac{W^2}{4\pi\lambda^2} \right)^N 
\exp\left[ -\pi\Gamma(\sigma')^2 W d \right] Q_N ,
\end{equation}
where
\begin{equation}
Q_N = \frac{1}{N!} \int_{\Lambda} \prod_{j=1}^N [{\rm d}^2 z_j\, w_{\rm ren}(x_j)]
\prod_{j<k} \left\vert {\rm e}^{-\frac{2\pi}{W}z_j} - {\rm e}^{-\frac{2\pi}{W}z_j}
\right\vert^{\Gamma} 
\end{equation}
and
\begin{eqnarray}
w_{\rm ren}(x) & = & \frac{4\pi}{W^2} \epsilon(x) \exp\left[ 
\pi\Gamma(\sigma'-\sigma)x + \frac{\pi\Gamma}{W} (N-1) x \right] \nonumber \\
& = & \frac{4\pi}{W^2} \epsilon(x) \exp\left[ 
\pi\Gamma\left( 2 \sigma'- \frac{1}{W} \right) x \right] \label{onebody}
\end{eqnarray}
is the renormalized one-body Boltzmann factor.

For $\Gamma=2\gamma$ ($\gamma$ a positive integer), the technique of 
anticommuting variables \cite{Samaj11b,Samaj95,Samaj04} allows us to 
express $Q_N$ as the integral over Grassman variables with the action 
that couples certain composite operators by interaction strengths
\begin{equation}
w_{jk} = \int_{\Lambda} {\rm d}^2 z\, w_{\rm ren}(x) 
\exp\left( -\frac{2\pi}{W} jz\right) 
\exp\left( -\frac{2\pi}{W} k \bar{z} \right) 
\end{equation}
$[j,k=0,1,\ldots,\gamma(N-1)]$.
Due to the orthogonality relation
\begin{equation}
\int_0^W {\rm d}y\, \exp\left[ \frac{2\pi}{W} {\rm i}(k-j)y \right]
= W \delta_{jk} , 
\end{equation}
the interaction matrix becomes diagonal, $w_{jk} = w_j \delta_{jk}$ with
\begin{equation} \label{diagel}
w_j = W \int_0^d {\rm d}x\, w_{\rm ren}(x) 
\exp\left( -\frac{4\pi}{W}jx \right) 
\end{equation} 
$[j=0,1,\ldots,\gamma(N-1)]$.

At the special coupling $\Gamma=2$ $(\gamma=1)$, the composite operators 
become the standard anticommuting variables. 
Due to the diagonalized form of the action 
\begin{equation} \label{action}
S=\sum_{j=0}^{N-1} \xi_j w_j \psi_j ,
\end{equation}
we obtain the exact result
\begin{equation}
Q_N = \prod_{j=0}^{N-1} w_j . 
\end{equation}
Omitting in (\ref{partf}) the irrelevant $d$-independent prefactors, 
the free energy, defined by $-\beta F_N = \ln Z_N$, reads
\begin{equation}
-\beta F_N = -2 \pi (\sigma')^2 W d + \sum_{j=0}^{N-1} \ln w_j .
\end{equation}
The corresponding pressure $P_N$, i.e. the force between the plates
per unit length of one of the plates, is given by
\begin{equation}
\beta P_N = \frac{\partial}{\partial d} \left( \frac{-\beta F_N}{W} \right) .
\end{equation}
Explicitly,
\begin{equation} \label{pressure}
\beta P_N = - 2\pi (\sigma')^2 + \frac{1}{W} \sum_{j=0}^{N-1}
\frac{1}{w_j} \frac{\partial w_j}{\partial d} .
\end{equation}
Moreover, using the formalism of anticommuting variables, the density profile 
of particles is expressible as
\begin{equation} \label{density}
n(x) = w_{\rm ren}(x) \sum_{j=0}^{N-1} \langle \xi_j \psi_j \rangle 
\exp\left( -\frac{4\pi}{W}jx \right) ,
\end{equation}
where the correlators are given by
$\langle \xi_j\psi_j\rangle = 1/w_j$ $(j=0,\ldots,N-1)$
within the free-fermion action (\ref{action}). 

\section{Counter-ions between charged walls}
\label{sec:3}
We first consider the unconstrained model in Fig. \ref{fig:1} with 
the trivial $\epsilon$-function given by (\ref{epsunconstrained}).
At $\Gamma=2$, the renormalized one-body weight (\ref{onebody}) reads
\begin{equation}
w_{\rm ren}(x) = \frac{4\pi}{W^2} \exp\left[ 
2\pi\left( 2 \sigma'- \frac{1}{W} \right) x \right]
\end{equation}
and the diagonal interaction elements (\ref{diagel}) are given by
\begin{equation}
w_j = \frac{1}{j-W\sigma'+\frac{1}{2}} \left[ 1 -
{\rm e}^{- \frac{4\pi d}{W} \left( j-W\sigma'+\frac{1}{2} \right)} 
\right]
\end{equation}
$(j=0,1,\ldots,N-1)$.
The pressure (\ref{pressure}) is then expressible as
\begin{eqnarray}
\beta P_N & = & -2\pi(\sigma')^2 + \frac{4\pi}{W^2} \sum_{j=0}^{N-1}
\frac{j-W\sigma'+\frac{1}{2}}{1 -
{\rm e}^{- \frac{4\pi d}{W} \left( j-W\sigma'+\frac{1}{2} \right)}} \nonumber \\
& & \times {\rm e}^{- \frac{4\pi d}{W} \left( j-W\sigma'+\frac{1}{2} \right)}
\end{eqnarray}
and the density profile (\ref{density}) as
\begin{equation} \label{densityy}
n(x) = \frac{4\pi}{W^2} \sum_{j=0}^{N-1} \frac{j-W\sigma'+\frac{1}{2}}{1 -
{\rm e}^{- \frac{4\pi d}{W} \left( j-W\sigma'+\frac{1}{2} \right)}}
{\rm e}^{- \frac{4\pi x}{W} \left( j-W\sigma'+\frac{1}{2} \right)} .
\end{equation}
The particle number density has the correct dimension [length]$^{-2}$.
It can be readily shown that 
\begin{equation} \label{pressure1}
\beta P_N = n(d) - 2\pi (\sigma')^2 .
\end{equation}
Simultaneously, Eq. (\ref{densityy}) implies the exact relation
\begin{equation}
n(0) - n(d) = 2\pi \left[ \sigma^2-(\sigma')^2 \right]
\end{equation}
and we can write
\begin{equation} \label{pressure2}
\beta P_N = n(0) - 2\pi \sigma^2 .
\end{equation}
Relations (\ref{pressure1}) and (\ref{pressure2}) correspond to the contact 
theorem valid for rectilinear (planar or line) wall surfaces 
\cite{Henderson78,Carnie81}.
It is interesting that the contact theorem has the same form for
arbitrary, finite or infinite, circumference $W$. 

Our next aim is to perform the continuum of the above formulas in the 
thermodynamic limit $N,W\to\infty$, at the fixed ratio $N/W=\sigma+\sigma'$.
We choose $t=(j-W\sigma'+\frac{1}{2})/N$ as the continuous variable.
For the density profile (\ref{densityy}), one obtains
\begin{eqnarray}
n(x) & = & 4\pi \left( \frac{N}{W} \right)^2 
\int_{-\frac{\sigma'}{\sigma+\sigma'}}^{\frac{\sigma}{\sigma+\sigma'}} {\rm d}t\, 
t \frac{{\rm e}^{-4\pi(\sigma+\sigma')t x}}{1-{\rm e}^{-4\pi(\sigma+\sigma')t d}}
\nonumber \\ & = & \frac{1}{\pi d^2} \int_{-2\pi\sigma' d}^{2\pi\sigma d} {\rm d}s\, s
\frac{{\rm e}^{-2 s x/d}}{1-{\rm e}^{-2s}} .
\end{eqnarray}
The pressure, given by
\begin{equation}
\beta P = \frac{1}{2} \left[ n(0) + n(d) \right] - 
\pi \left[ \sigma^2 +(\sigma')^2 \right] ,
\end{equation}
is found after some algebra in a symmetrized form
\begin{equation} \label{press1}
\beta P = \frac{1}{2\pi d^2} \left( 
\int_0^{2\pi\sigma d} + \int_0^{2\pi\sigma' d} \right) 
{\rm d}s\, \frac{s}{\sinh s} {\rm e}^{-s} . 
\end{equation}
In the symmetric case $\sigma=\sigma'$, we reproduce the previously 
obtained formula (4.40) of ref. \cite{Samaj11b}.

We want to analyze the dependence of the pressure on the distance $d$ 
between the charged lines.
Using the substitution $s=d t$ in (\ref{press1}), the differentiation 
with respect to $d$ implies
\begin{equation} \label{monpress1}
\frac{\partial}{\partial d} \beta P = - \frac{1}{2\pi}
\int_{-2\pi\sigma'}^{2\pi\sigma} {\rm d}t\, \left[ \frac{t}{\sinh(d t)} \right]^2
< 0 ,
\end{equation}
where the integral is positive with regard to the restriction 
$-\sigma'<\sigma$.
This means that $\beta P$ decays monotonously as the function of 
distance $d$.

In the limit of small distances, we see from (\ref{press1}) that the pressure
\begin{equation}
\beta P \mathop{\sim}_{d\to 0} \frac{\sigma+\sigma'}{d} > 0 
\end{equation} 
exhibits the singularity  and is positive, i.e. there is a repulsion between 
the charged lines. 
This small $d$ divergence is expected, and simply stems from the entropy cost 
of confining the mobile ions in a narrow slab. 
Quite clearly then, we may anticipate that the constrained model will not 
exhibit the same $1/d$ divergence, since approaching the two charged 
lines does not restrict the configurational space of mobile ions.

\begin{figure}[tbp]
\begin{center}
\includegraphics[width=0.8\textwidth,clip]{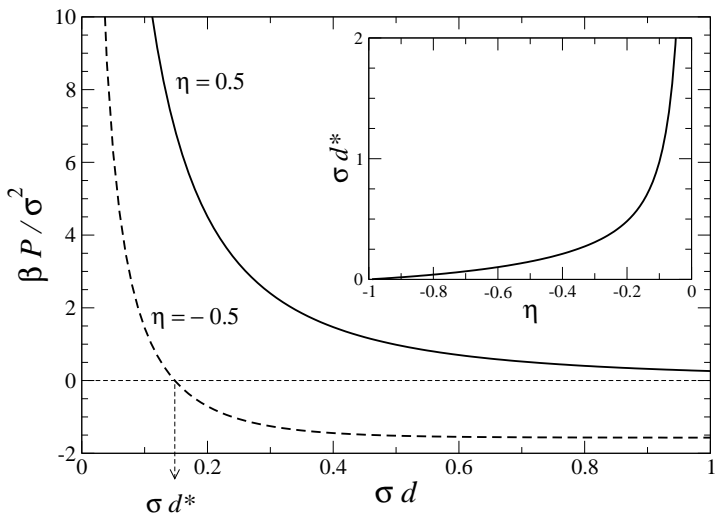}
\caption{Unconstrained model:
The dimensionless pressure $\beta P/\sigma^2$ vs. the dimensionless 
distance $\sigma d$ for two values of the asymmetry parameter 
$\eta=1/2$ (solid line) and $\eta=-1/2$ (dashed line). 
In both cases, the pressure exhibits a monotonous decrease from $\infty$ at 
$\sigma d\to 0$ to: 0 for $\eta=1/2$ and the capacitor limit $-2\pi\eta^2$ 
for $\eta=-1/2$ at $\sigma d\to\infty$.
In the latter case, the pressure equals to zero at the equilibrium distance 
$\sigma d^*$ whose dependence on the negative  $\eta$
is pictured in the inset.}
\label{fig:3}
\end{center}
\end{figure}

The asymptotic large-distance limit of $\beta P$ depends on the sign
of $\sigma'$.
\begin{itemize}
\item
If $\sigma'>0$ (like-charged lines), we have the universal result
\begin{equation} \label{sigmapos}
\beta P \mathop{\sim}_{d\to\infty} \frac{1}{\pi d^2} 
\int_0^{\infty} {\rm d}s\, \frac{s}{\sinh s} {\rm e}^{-s}
= \frac{\pi}{12} \frac{1}{d^2} ,
\end{equation}
independent of the amplitudes of $\sigma$ and $\sigma'$ 
\footnote{Note that whenever the pressure is of the form 
$\beta P \propto 1/d^2$, it cannot depend on the absolute values of 
$\sigma$ and $\sigma'$, but only on the ratio $\eta=\sigma/\sigma'$,
for dimensional reasons. In the symmetric case, it is thus 
$\sigma$ independent.}.
It coincides with the previous finding (\ref{Gamma2}) for symmetrically
charged lines; the same phenomenon is observed in the mean-field limit where 
the formula (\ref{meanfield}) takes place for any $\sigma'>0$ \cite{Kanduc08}. 
The pressure goes to zero at $d\to\infty$ from above, in agreement 
with its monotonous decrease property from $+\infty$ at $d\to 0$.
Another consequence is that the pressure is positive (repulsive) for every 
$d$; the value of the coupling $\Gamma=2$ is not large enough
to obtain attraction between the like-charged lines.
The plot of the dimensionless pressure $\beta P/\sigma^2$ vs. the
dimensionless distance $\sigma d$ for $\eta=1/2$ is pictured 
in Fig. \ref{fig:3} by the solid line.
The regions close to the two charged lines decouple from one another
in the limit $d\to\infty$, the densities of particles at the lines
$n(0)=2\pi\sigma^2$ and $n(d)=2\pi(\sigma')^2$ being equivalent to those 
for semi-infinite geometries. 
\item
For a neutral line at $x=d$, $\sigma'=0$, we have
\begin{equation}
\beta P \mathop{\sim}_{d\to\infty} \frac{1}{2\pi d^2} 
\int_0^{\infty} {\rm d}s\, \frac{s}{\sinh s} {\rm e}^{-s}
= \frac{\pi}{24} \frac{1}{d^2} .
\end{equation} 
As before, the pressure is always positive.
Comparing this result with (\ref{sigmapos}) we see that the neutrality
of the line at $x=d$ diminishes the asymptotic pressure by $1/2$. 
In the mean-field limit, the diminishing factor is equal to $1/4$ 
\cite{Kanduc08}; that factor $1/4$ is obtained by redefining $d\to 2d$ which
was argued to be related to the decomposition of a system with an asymmetry
parameter $\eta>0$ into two halves each with an effective $\eta=0$.
Such an argument, which discards fluctuations, no longer applies to 
the coupling constant $\Gamma=2$. 
\item
If $\sigma'<0$ (oppositely charged lines), we reexpress the second integral 
on the rhs of (\ref{press1}) by using the substitution $s\to -s$ and
afterwards writing ${\rm e}^s = {\rm e}^s - {\rm e}^{-s} + {\rm e}^{-s}$,
with the result
$$- (2\pi\sigma' d)^2 - \int_0^{-2\pi\sigma' d} {\rm d}s\, 
\frac{s}{\sinh s} {\rm e}^{-s} . $$
Thus, at large distances, the pressure does not vanish,
\begin{equation} \label{capacitor1}
\beta P \mathop{\sim}_{d\to\infty} = -2\pi (\sigma')^2
\end{equation} 
with an exponentially decaying correction.
This behaviour is not surprising.
Since the sign of the particle charge is the same as that of the line at 
$x=d$, particles cannot neutralize and are repelled from this line.
For sufficiently distant lines we have $n(0)\simeq 2\pi[\sigma^2-(\sigma')^2]$
and $n(d)\simeq 0$, i.e. all counter-ions stay in the neighbourhood
of line at $x=0$ and compensate partially its line charge density
$\sigma e$ to $\vert\sigma'\vert e$.
We are left with a capacitor of two lines with opposite charge densities
$\pm \sigma' e$ for which the pressure is negative (attractive) and
given just by Eq. (\ref{capacitor1}). 
The same scenario was discussed in \cite{PaTr11}, and occurs in 
the mean-field treatment \cite{Kanduc08}.
The dependence of $\beta P/\sigma^2$ on $\sigma d$ for $\eta=-1/2$ is 
represented in Fig. \ref{fig:3} by the dashed line.
Since $\beta P$ decays monotonously from $+\infty$ at $d\to 0$ to
the negative number $-2\pi (\sigma')^2$ at $d\to\infty$, there exists just
one equilibrium distance $d^*$ at which $\beta P(d^*) = 0$.
Thus, oppositely charged lines repel one another for distances $d<d^*$ 
while there is an attraction for $d>d^*$. 
The dependence of $\sigma d^*$ on the negative asymmetry $\eta$
is presented in the inset of Fig. \ref{fig:3}.
\end{itemize}

\section{Counter-ions at charged walls}
\label{sec:4}
The constrained model in Fig. \ref{fig:2} possesses a non-trivial 
$\epsilon$-function (\ref{epsconstrained}).
At $\Gamma=2$, the renormalized one-body weight (\ref{onebody}) reads
\begin{equation}
w_{\rm ren}(x) = \frac{4\pi\lambda}{W^2} \left[ \delta(x)+\delta(x-d) \right] 
\exp\left[ 2\pi\left( 2 \sigma'- \frac{1}{W} \right) x \right]
\end{equation}
and the diagonal interaction elements (\ref{diagel}) are given by
\begin{equation}
w_j = \frac{4\pi\lambda}{W}\left[ 1 +
{\rm e}^{- \frac{4\pi d}{W} \left( j-W\sigma'+\frac{1}{2} \right)} \right]
\end{equation}
$(j=0,1,\ldots,N-1)$.
The pressure (\ref{pressure}) is given by
\begin{eqnarray}
\beta P_N & = & -2\pi(\sigma')^2 - \frac{4\pi}{W^2} \sum_{j=0}^{N-1}
\frac{j-W\sigma'+\frac{1}{2}}{1 +
{\rm e}^{- \frac{4\pi d}{W} \left( j-W\sigma'+\frac{1}{2} \right)}} \nonumber \\
& & \times {\rm e}^{- \frac{4\pi d}{W} \left( j-W\sigma'+\frac{1}{2} \right)} ,
\end{eqnarray}
which is of course independent of $\lambda$.
It is easy to show that this relation is equivalent to the following one
\begin{equation}
\beta P_N = -2\pi\sigma^2 + \frac{4\pi}{W^2} \sum_{j=0}^{N-1}
\frac{j-W\sigma'+\frac{1}{2}}{1 +
{\rm e}^{- \frac{4\pi d}{W} \left( j-W\sigma'+\frac{1}{2} \right)}} ,
\end{equation}
so that we can write
\begin{eqnarray}
\beta P_N & = & -\pi\left[ \sigma^2+(\sigma')^2 \right] 
+ \frac{2\pi}{W^2} \sum_{j=0}^{N-1} \left( j-W\sigma'+\frac{1}{2} \right)
\nonumber \\ & & \times
\tanh\left[ \frac{2\pi d}{W} \left( j-W\sigma'+\frac{1}{2} \right) \right] .
\end{eqnarray}
The continuum procedure, analogous to that for the unconstrained
model, leads to
\begin{equation} \label{press2}
\beta P = - \frac{1}{2\pi d^2} \left( 
\int_0^{2\pi\sigma d} + \int_0^{2\pi\sigma' d} \right)
{\rm d}s\, \frac{s}{\cosh s} {\rm e}^{-s} . 
\end{equation}  

\begin{figure}[tbp]
\begin{center}
\includegraphics[width=0.8\textwidth,clip]{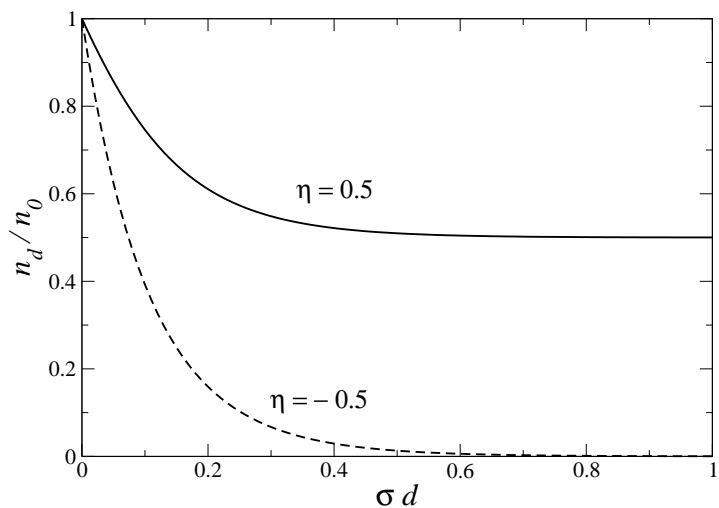}
\caption{Constrained model:
The ratio of the line particle densities $n_d/n_0$ vs. distance
$\sigma d$ for the asymmetry $\eta=1/2$ (solid line) and $\eta=-1/2$ 
(dashed line). 
In both cases, the ratio decreases monotonously from 1 at $\sigma d\to 0$
to: the electroneutrality limit $0.5$ for $\eta=1/2$ and 0 for $\eta=-1/2$ 
at $\sigma d\to\infty$.}
\label{fig:4}
\end{center}
\end{figure}

The density profile (\ref{density}) has the form
\begin{equation}
n(x) = n_0 \delta(x) + n_d \delta(x-d) ,
\end{equation}
where
\begin{equation} \label{n0}
n_0 = \frac{1}{W} \sum_{j=0}^{N-1} \frac{1}{1 +
{\rm e}^{- \frac{4\pi d}{W} \left( j-W\sigma'+\frac{1}{2} \right)}} 
\end{equation}
and
\begin{equation} \label{nd}
n_d = \frac{1}{W} \sum_{j=0}^{N-1} \frac{1}{1 +
{\rm e}^{- \frac{4\pi d}{W} \left( j-W\sigma'+\frac{1}{2} \right)}} 
{\rm e}^{- \frac{4\pi d}{W} \left( j-W\sigma'+\frac{1}{2} \right)} .
\end{equation}
The prefactors $n_0$ and $n_d$ to the Dirac delta functions are the line 
particle densities of dimension [length]$^{-1}$ along the $y$-axis 
at $x=0$ and $x=d$, respectively.
They are constrained by the obvious electroneutrality condition
\begin{equation}
n_0 + n_d = \frac{N}{W} = \sigma+\sigma' .
\end{equation}
After taking the continuum limits of (\ref{n0}) and (\ref{nd}), we get
\begin{eqnarray}
n_0 & = & \sigma + \frac{1}{4\pi d} \ln \left( 
\frac{1+{\rm e}^{-4\pi\sigma d}}{1+{\rm e}^{-4\pi\sigma' d}} \right) , \\
n_d & = & \sigma' + \frac{1}{4\pi d} \ln \left( 
\frac{1+{\rm e}^{-4\pi\sigma' d}}{1+{\rm e}^{-4\pi\sigma d}} \right) .
\end{eqnarray}
It holds that $0\le n_d\le n_0$.
Note that for any finite distance $d$, (and symmetric $\sigma=\sigma'$ case 
excluded), the net (fixed background plus mobile particle) charge density on 
each of the two lines is nonzero: $(\sigma-n_0)e\ne 0$ and 
$(\sigma'-n_d)e = -(\sigma-n_0)e\ne 0$.
The neutrality of the two lines is attained at asymptotically large $d$ 
provided that $\sigma'\ge 0$.
If $\sigma'<0$, we find that $n_0=\sigma+\sigma'$ and $n_d=0$, i.e. the system
corresponds to a capacitor of two lines with opposite charge densities
$\pm\sigma' e$. 
The ratio $n_d/n_0$ as a function of the dimensionless distance $\sigma d$
is drawn for two values of the asymmetry parameter $\eta=1/2$ (solid line)
and $\eta=-1/2$ (dashed line) in Fig. \ref{fig:4}.
It is seen that $n_d/n_0\to 1$ for $\sigma d\to 0$ for every positive or 
negative value of $\eta$. 
This is an entropy driven phenomenon, since a particle's localization on one 
or the other line is at small $d$ inconsequential from the energetic point 
of view.
For $\eta=1/2$, the ratio $n_d/n_0$ tends to the neutrality value $1/2$
at $\sigma d\to\infty$.
On the other hand, for $\eta=-1/2$, the ratio $n_d/n_0=0$ at 
$\sigma d\to\infty$, in agreement with the above discussion.
It is interesting to note that for negative $\eta$ the value of $n_d$ is 
strictly positive for finite values of $\sigma d$: although the energy is 
increased by putting particles to the likely charged line, the entropy 
increase on this less populated line favors a non-vanishing occupation 
by particles.

It can be shown from the expression of the pressure (\ref{press2}) that
\begin{equation} \label{monpress2}
\frac{\partial}{\partial d} \beta P = \frac{1}{2\pi}
\int_{-2\pi\sigma'}^{2\pi\sigma} {\rm d}t\, \left[ \frac{t}{\cosh(d t)} \right]^2
> 0 ,
\end{equation}
i.e. the pressure is a monotonously increasing function of $d$.
In the limit $d\to 0$, $\beta P$ is negative:
\begin{equation} \label{smalld}
\beta P \mathop{\sim}_{d\to 0} = - \pi \left[ \sigma^2 + (\sigma')^2 \right] ,
\end{equation} 
and, as expected, does not exhibit any singularity.

\begin{figure}[tbp]
\begin{center}
\includegraphics[width=0.8\textwidth,clip]{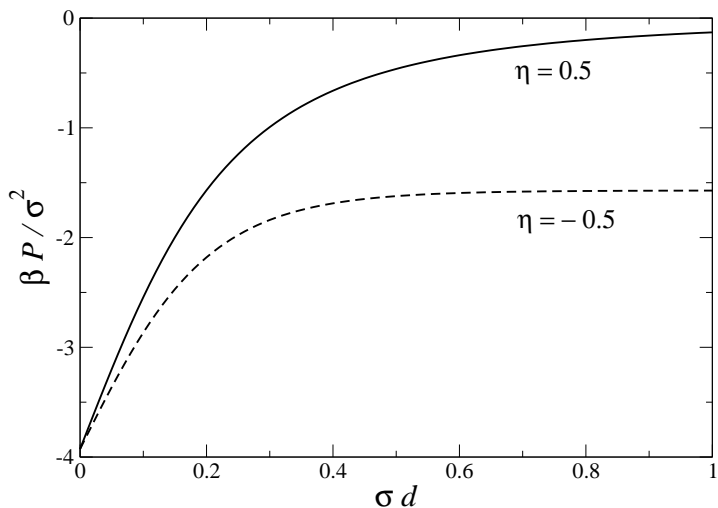}
\caption{Constrained model:
Dimensionless pressure $\beta P/\sigma^2$ as a function of 
dimensionless distance $\sigma d$ for the asymmetry $\eta=1/2$ (solid line) 
and $\eta=-1/2$ (dashed line). 
In both cases, the pressure exhibits a monotonous increase from 
$-\pi(1+\eta^2)$ at $\sigma d\to 0$ to: 0 for $\eta=1/2$ and the capacitor 
limit $-2\pi\eta^2$ for $\eta=-1/2$ at $\sigma d\to\infty$.}
\label{fig:5}
\end{center}
\end{figure}

At asymptotically large distances, the sign of $\sigma'$ is important.
\begin{itemize}
\item
If $\sigma'>0$, one obtains the universal formula
\begin{equation} \label{univpres}
\beta P \mathop{\sim}_{d\to\infty} - \frac{1}{\pi d^2} 
\int_0^{\infty} {\rm d}s\, \frac{s}{\cosh s} {\rm e}^{-s}
= - \frac{\pi}{24} \frac{1}{d^2} .
\end{equation}
This asymptotic behaviour coincides with that of the van der Waals-Casimir
for two neutral 2D conductors (\ref{Casimir2D}); since $n_0=\sigma$ and 
$n_d=\sigma'$ at infinite distance, each of the two lines is indeed neutral 
and the force between the lines results from internal charge fluctuations 
inside each of the lines.  
The pressure goes to zero at $d\to\infty$ from below, which agrees 
with its property of the monotonous increase from the negative value 
(\ref{smalld}) at $d\to 0$. 
The pressure is always negative (attractive) for the couple of
like-charged lines.
The dimensionless pressure plot for $\eta=1/2$ is presented in Fig. 
\ref{fig:5} by the solid line.
\item
For $\sigma'=0$, we find that
\begin{equation}
\beta P \mathop{\sim}_{d\to\infty} - \frac{1}{2\pi d^2} 
\int_0^{\infty} {\rm d}s\, \frac{s}{\sinh s} {\rm e}^{-s}
= - \frac{\pi}{48} \frac{1}{d^2} ,
\end{equation} 
which is one half of the previous result (\ref{univpres}).
As before, the pressure is negative.
\item
If $\sigma'<0$, we reexpress the second integral on the rhs of 
(\ref{press2}) as
$$(2\pi\sigma' d)^2 - \int_0^{-2\pi\sigma' d} {\rm d}s\, 
\frac{s}{\cosh s} {\rm e}^{-s}$$
and arrive at the capacitor result
\begin{equation} \label{capacitor2}
\beta P \mathop{\sim}_{d\to\infty} = -2\pi (\sigma')^2 .
\end{equation} 
Since the monotonously increasing $\beta P$ interpolates between
two negative values as $d$ goes from 0 to $\infty$, the pressure 
is always negative.
For the asymmetry parameter $\eta=-1/2$, $\beta P/\sigma^2$ as the
function of $\sigma d$ is pictured in Fig. \ref{fig:5} by the dashed line.
\end{itemize}

\section{Conclusion}
\label{sec:concl}
In this paper, we investigated thermal equilibrium of 2D Coulomb systems 
composed of two parallel asymmetrically charged lines at distance $d$,
neutralized by ``counter-ions only''.
Two versions of the model were considered: in the unconstrained version, 
counter-ions move freely in the region between the lines, while in 
the constrained version, counter-ions stick to the charged lines.
For the exactly solvable coupling constant $\Gamma=2$, we analyzed 
the dependence of the pressure on $d$, especially its sign and 
the asymptotic large-$d$ behaviour.

The two models exhibit some common features.  
The expressions for the pressure (\ref{press1}) and (\ref{press2}) consist
of the sum of two decoupled contributions, one from each line. 
Such property does not occur in the weak-coupling limit and is probably 
related to the free-fermion nature of the $\Gamma=2$ coupling. 
According to the derivative formulas (\ref{monpress1}) and (\ref{monpress2}),
the pressure is always a monotonous function of $d$, decreasing for
the unconstrained model, and increasing for its constrained counterpart.
A similarly increasing pressure is found for bilayers, frozen in their
ground state (with thus a diverging coupling constant) \cite{SaTr12}.

For likely charged lines, the pressure goes to 0 at large $d$ without 
changing its sign: it is always positive (repulsive) for the unconstrained 
model and negative (attractive) for the constrained model.
The large-distance asymptotic behaviour of the pressure for the unconstrained 
model (\ref{sigmapos}) is universal, however, the prefactor to $1/d^2$
{\em is not} consistent with the mean-field prediction (\ref{meanfield})
but has to be renormalized by a temperature-dependent function.
This remark on the failure of mean-field at asymptotic distances,
corroborates previous 2D results obtained for symmetric lines \cite{Samaj11b}.
It goes against common expectation that mean-field should hold at large $d$ 
\cite{Shklovskii,Levin09}, the underlying argument being that the small 
density of ions far from the plate may effectively drive the system into 
a weakly-coupled regime, in the corresponding distance range.
However, it should be kept in mind that Coulomb potential $v({\bf r})$ 
is logarithmic in two dimensions and hence scale-free, so that the Coulombic 
coupling does not depend on the density: it is always $\beta e^2$.
Hence, a 2D system, unlike its higher dimensional counterparts, is nowhere 
weakly coupled when $\Gamma>1$, and it therefore does not come as a surprise
that mean-field breaks down in 2D, even at large distances. 
Conversely, in dimension $\nu=3$, the standard argument on the asymptotic 
validity of mean-field may well apply.
We note that it is backed up by recent accurate Monte Carlo results 
\cite{Mallarino}.
As concerns the constrained model, the net charge density on each of 
the two lines is nonzero for every finite distance $d$ and vanishes only
at $d\to\infty$.
The universal $d\to\infty$ behavior of the pressure (\ref{univpres}) 
coincides with that of the van der Waals-Casimir for two neutral 2D 
conductors (\ref{Casimir2D}), without any need for temperature renormalization.

For oppositely charged lines, the pressure at asymptotically large $d$
corresponds to a capacitor in both unconstrained (\ref{capacitor1}) and 
constrained (\ref{capacitor2}) cases.
This fact has an important impact especially on the unconstrained model 
which exhibits at a special dimensionless distance $\sigma d^*$ a transition 
from the repulsion regime for $d<d^*$ to the attraction regime for $d>d^*$;
for the dependence of $\sigma d^*$ on the negative asymmetry 
$\eta\equiv \sigma'/\sigma$ see the inset of Fig. \ref{fig:3}. 

Finally, investigating the two present models under stronger correlations
(e.g. $\Gamma=4$, or 6) and also ultimately in the strongly coupled limit 
($\Gamma \to \infty$), provides interesting venues for future work.

\begin{acknowledgement}
The support received from Grant VEGA No. 2/0049/12 is acknowledged.
\end{acknowledgement}

\end{document}